\documentclass[a4paper,12pt]{article}
\usepackage[utf8]{inputenc}
\usepackage[english]{babel}
\usepackage{amssymb}
\usepackage{amsmath}
\usepackage{epsfig}
\usepackage{bm}
\usepackage{graphics}\textwidth=17cm
\def\be{\begin{equation}}
\def\ee{\end{equation}}

\def\deg{{\rm deg}}

\def\sign{{\rm sign}}
\newtheorem{thrm}{\bf Theorem}
\newtheorem{lmm}{\bf Lemma}
\newtheorem{rmk}{\bf Remark}
\newtheorem{dfn}{\bf Definition}
\def\bl{\rule[-1mm]{2.4mm}{2.4mm}}
\def\Arg{{\rm Arg}}

\graphicspath{{filters/Filter-strategy/img/}}

\begin{document}

\title{Projective view at Optimization Problem for\\ Multiband Filter}
\author{Andrei Bogatyr\"ev
\thanks{Supported by RSF grant 16-11-10349P}
}
\date{}
\maketitle

\abstract{The best uniform rational approximation of the \emph{sign} function on two
intervals separated by zero was explicitly found by E.I. Zolotar\"ev in 1877. This optimization problem 
is the initial step in the staircase of the so called approximation problems for multiband filters 
which are of great importance for electrical engineering. 
We show that known in the literature optimality criterion for this problem may be contradictory since it does 
not take into account the projective invariance of the problem. 
We propose a new consistently projective formulation of this problem 
and give a constructive optimality criterion for it.}


\section*{Introduction}
The design of modern electronic devices involves a lot of deep and sophisticated mathematics.
In particular, synthesis of multiband analogue, digital and microwave electrical filters despite their 
diverse physical implementation  is based on the same uniform rational approximation problem which 
is a generalization of third and fourth Zolotar\"ev problems. This problem attracted attention of many mathematicians
including such influential figures  as E.I.Zolotar\"ev, N.I.Achieser, E.Stiefel, A.A.Gonchar 
to say nothing of numerous electrical and electronic engineers
who are interested mainly in the algorithmic aspects of the problem. 

\begin{figure}
\includegraphics[width=\textwidth]{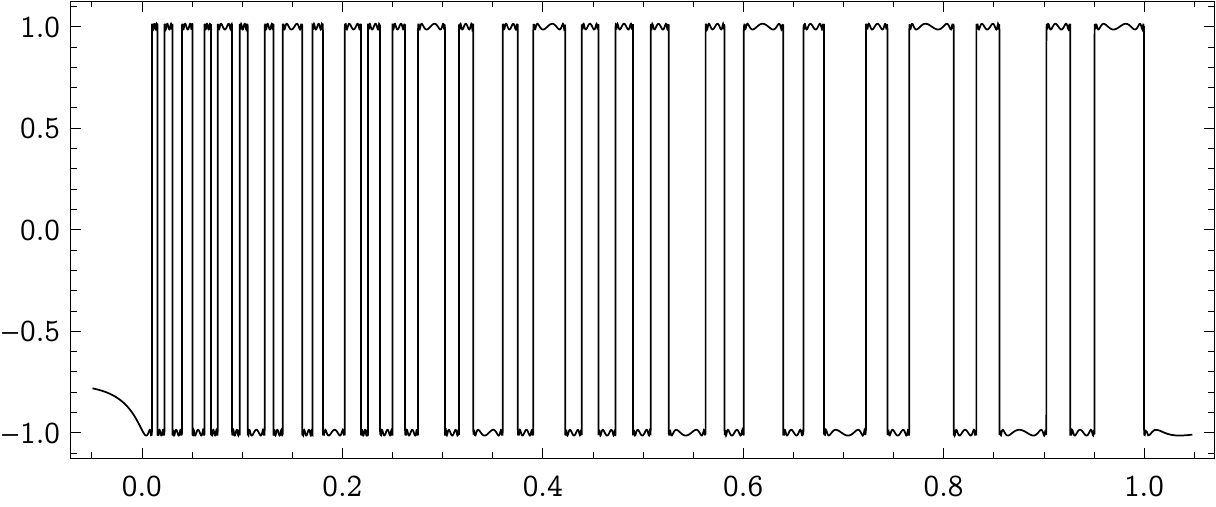}
\caption{The optimizer for the 55 workbands set $E$ obtained by the Ansatz method and computed by S.Lyamaev.}
\label{optimus} 
\end{figure}

Roughly, the problem consists in the best uniform approximation of the 
two-valued (indicator) function defined on the prescribed  passbands and stopbands -- the collection $E$ of disjoint segments on real frequency axis -- by a  given degree $n$ rational function. It was nearly immediately discovered that 
even in the simplest two-band classical variant the problem is multi-extremal, that is admits local optima. 
It is likely that Ed Stiefel \cite{Sti} with his pupil R.A.-R.Amer \cite{AS} were the first who decomposed the set of rational functions into 
topological classes which secured the uniqueness of the solution. Each of those solutions may be characterized by the alternation principle
-- or the equiripple property in terms of engineers -- which says that the graph of (the error function of the) solution looks like 
the sequence of sufficiently many ripples of constant amplitude, see Fig. \ref{optimus}.

With this property taken into account, the solution of the problem becomes in a sense very simple: you merely show a function with the requested  oscillatory behaviour. 
Several approaches to the numerical solution of optimization problem for multiband 
electrical filter are based on the alternation principle. 

Say, Remez-type algorithms specially designed for uniform polynomial and rational approximation \cite{Remez,Veidinger,Fuchs,Sti2}
iteratively build the necessary alternation set for the error function of approximation.
For more traditional gradient descent methods the appearance of the alternation is a
signal to terminate iterations: equiripple property serves as a certificate for the optimizer. 

Another approach for the solution is the recently elaborated Ansatz method \cite{BGL,Bbook}, 
the development of the classical approach. An explicit analytical formula for the solution was proposed
by  E.I.Zolotarev \cite{Zol} in case of two bands,  later extended by E.Stiefel \cite{Sti} to three bands and 
independently by this author \cite{B10} for arbitrary number of bands. 
This formula -- the elliptic sine of an abelian integral -- generalizes the representation for Zolotar\"ev fractions and contains unknown parameters, both continuous and discrete, which have to be evaluated given the input data of the problem. 
The number of Ansatz parameters is usually  much less than the number of optimization variables, which makes up the heart of this approach.
The Ansatz automatically obeys the equiripple criterion, however it becomes a rational function for only 
specially chosen sets of parameters.

Careful formulation of optimality criterion is very important, however up to now this was not done fairly well for the 
solution with defect, when its algebraic degree drops.  Known criterion proposed in \cite{Malo} does not 
endure the projective transformations of the set $E$ of the filter workbands, the important property of this optimization problem.
Also it does not contain any information of the above mentioned topological classes which guarantee the uniqueness of the optimizer.
We propose a modification of Malozemov's criterion for the defect $d$ optimizer $R(x)$ in the fixed topological class of degree $n$ rational functions: 
$$
Alt(R)\ge 2n+2-d-\Sigma^0+\Sigma^1,
$$
where $Alt(R)$ is the number of cyclic alternation points on the set $E$ of workbands of the filter,
$\Sigma^0$, $\Sigma^1$ are nonnegative integer indexes which depend on the solution $R$, the topological class and vanish in the case
of full degree solution. We emphasize that both sides of the inequality are  now even numbers.

\section{Optimization problem for multiband filter}
Suppose a finite collection $E$ of disjoint closed segments of real axis $\mathbb{R}$ is given.
The set has a meaning of frequency bands and is decomposed into two parts: $E=E^+\cup E^-$
which are called the passbands $E^+$ and the stopbands $E^-$. Both subsets $E^\pm$ are non empty.
Optimization problem for electrical filter has   several equivalent settings \cite{Cauer, AS, Akh, Zol, Malo}.

\subsection{Three settings}\label{4Set}
In each of listed below cases we minimize certain quantity among real rational functions $R(x)$ of 
bounded degree $\deg R\le n$ (being the maximum of the  degrees of the numerator and the denominator of a fraction). 
The goal function may be one of the following. 

\subsubsection{Minimal deviation}\label{minDev}
$$\frac{\max_{x\in E^+}|R(x)|}{\min_{x\in E^-}|R(x)|}\longrightarrow \min =:\theta^2\le1.$$

\subsubsection{Third Zolotar\"ev problem}\label{Zolo3} 
Minimize $\theta$ under the condition that there exists a real rational function $R(x)$, 
$~\deg R\le n$, with the restrictions
	$$\min_{x\in E_-}|R(x)|\ge\theta^{-1},\quad \max_{x\in E_+}|R(x)|\le\theta.$$

\subsubsection{Fourth Zolotar\"ev problem} \label{Zolo4}
Define the indicator function $S_E(x)=\pm1$ when $x\in E^\pm$. Find the best uniform rational
approximation $R(x)$  of $S_E(x)$ of the given degree: 
  $$ 
 ||R-S_E||_{C(E)} := \max\limits_{x \in E}|R(x) - S_E(x)| \to \min =: \mu.
  $$	
  
It is a good exercise to show that all three settings are equivalent and in particular the value of $\theta$ is the same for the first two settings and $1/\mu=(\theta+1/\theta)/2$ for the third one.

\subsection{Historical remarks}
Setting \ref{minDev} appears in the papers \cite{Sti} by E.Stiefel (1961) and  \cite{AS} by R.A.-R.Amer, H.R.Schwarz (1964). 
A.A.Gonchar \cite{Gonchar} found the asymptotic of the minimal deviation value $\theta^2$ for large degree $n$ in terms of capacity of the condenser with plates $E^+$, $E^-$. Setting \ref{Zolo3} appears after suitable normalization of the rational function 
in \ref{minDev} and essentially coincides with the third Zolotar\"ev problem \cite{Zol}. Setting \ref{Zolo4} corresponds to the (extended) fourth  Zolotar\"ev problem \cite{Zol} and was studied by N.I.Akhiezer \cite{Akh} (1929). In \cite{Zol,Akh} they noticed that already in the classical Zolotar\"ev case with one pass- and one stop- band,  the minimizing function is not unique. This phenomenon was observed in the Stiefel's paper \cite{Sti} and fully explained in the  dissertation of his pupil R.-A.R.Amer \cite{AS} who decomposed the space of rational functions of bounded deviation (defined in the left hand side of formula in \ref{minDev}) into classes. Namely, the competing rational functions have no poles at the passbands and no zeros at the stopbands, hence  the sign of the polynomial in the numerator of the fraction on each stopband as well as  the sign of denominator polynomial on each passband is fixed. Then in the closure of each nonempty class there is a unique minimum. All mentioned authors detected  that (local) optimal functions are characterized by \emph{alternation} (or \emph{equiripple} in terms of electrical engineers) property. For instance, in the fourth Zolotar\"ev problem the approximation error $\delta(x):=R(x)-S_E(x)$ of degree $n$ minimizer has $2n+2$ \emph{alternation} points $a_s\in E$ where  $\delta(a_s)=\pm||\delta||_{C(E)}$ with consecutive change of sign.

\subsection{Equiripple criterion}
V.N.Malozemov in 1979  suggested \cite{Malo} a criterion for the defect $d$ solution (i.e. of exact degree $n-d$) 
which should possess at least $2n+2-d$ alternation points
to be an optimizer. This however may contradict the projective invariance of the problem. The latter property consists in the 1-1 correspondence between the solutions of the problems with the bands $E$ related by a projective (= linear-fractional) transformation. The number of alternation points will depend on the starting point of their counting once $d$ is odd. Also, the criterion does not take into consideration any information about Stiefel classes, which means that a minimizer with defect will be the minimizer for all classes whose closure contain it.

\begin{figure}[h]
\begin{picture}(120, 40)(-15,0)
 \multiput(0,0)(0,10){5}{\line(1,0){120}}
 \put(122,-2){$T_3=-1$}
 \put(122,8){$T_3=-\frac12$}
 \put(122,38){$T_3=1$}
 \put(122,28){$T_3=\frac12$}
 \qbezier(35,40)(45,40)(60,20)
 \qbezier(35,40)(23,40)(8,-5)
 \qbezier(85,0)(75,0)(60,20)
 \qbezier(85,0)(97,0)(112,45)
\put(10,20){\circle*{1}}
\put(-3,16){$-1=x_0$}

\put(13.5,20){\circle*{1}}
\put(12,22){$x_1$}

\put(23,20){\circle*{1}}
\put(22,17){$x_2$}

\put(35,20){\circle*{1}}
\put(33,16){$x_3$}

\put(52,20){\circle*{1}}
\put(51,17){$x_4$}

\put(68,20){\circle*{1}}
\put(67,22){$x_5$}

\put(85,20){\circle*{1}}
\put(83,22){$x_6$}

\put(97,20){\circle*{1}}
\put(96,22){$x_7$}

\put(107,20){\circle*{1}}
\put(105.5,17){$x_8$}

\put(110,20){\circle*{1}} 
\put(109,22){$x_9=1$}

\linethickness{.6mm}
\multiput(10,0)(3.5,0){2}{\line(0,1){10}}
\multiput(10,0)(0,10){2}{\line(1,0){3.5}}

\multiput(23,30)(29,0){2}{\line(0,1){10}}
\multiput(23,30)(0,10){2}{\line(1,0){29}}

\multiput(68,0)(29,0){2}{\line(0,1){10}}
\multiput(68,0)(0,10){2}{\line(1,0){29}}

\multiput(110,40)(-3.5,0){2}{\line(0,-1){10}}
\multiput(110,40)(0,-10){2}{\line(-1,0){3.5}} 
\end{picture}

\caption{Chebyshev polynomial $T_3(x)$ as a rational $n=4$ optimizer with defect $d=1$}
\label{T3}
\end{figure}
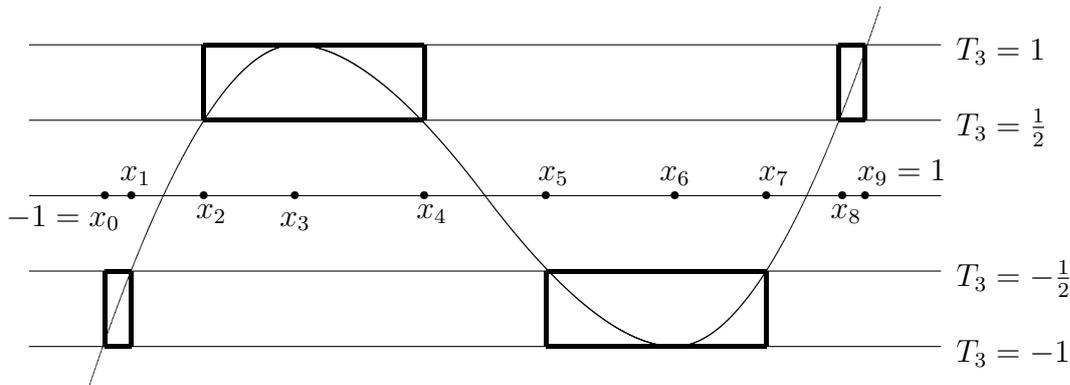

A simple counterexample to the criterion adapted for the setting \ref{Zolo4} is as follows (it can be transformed to any other above setting and to higher degree $n$). Degree three Chebyshev polynomial $T_3(x):=4x^3-3x$ approximates the function $\frac34 S_E(x)$ with the error $\mu:=\frac14$ on the set $E$ composed of bands $E^+=-E^-:=T_3^{-1}[\frac12,1]=[\cos(7\pi/9), \cos(5\pi/9)]\cup [\cos(\pi/9),1]$ . The error
function has 10 explicitly listed alternation points $x_j=-cos(j\frac\pi9)$, $j=0,\dots,9$
shown in the Fig. \ref{T3}. This is no doubt defect $d=1$ optimizer among degree $n=4$ rational functions (in the class where the function belongs to). The number of alternation points is slightly excessive so we can get rid of one of them, say $x_0=-1$, by pressing inside $E$ its very left endpoint. For the modified set $E^*$,  say $E^*:=E\setminus[-1,-0.99)$, same polynomial $T_3(x)$ will be again a (local) optimizer among degree $4$ rational functions by criterion of \cite{Malo} since it has 9 alternation points. We expect that the function $T_3\circ l$ will be an optimizer for the new set $E':=l^{-1}E^*$, say for $l(x):=1/x$  since the problems are obviously related. However, the same criterion says that $T_3\circ l$ will be \emph{not} an optimizer in any class of degree 4 functions as it admits only 8 alternation points: only one of the points $l(x_1)$, $l(x_9)$ may be included in the alternation set since the error of approximation is positive for both points.

Examples of this kind may be given for any (not too low) degree $n$ and the total number of bands $m\ge3$
with the explicit construction of equiripple rational functions from \cite{B10}.

\section{Projective setting}
Here we discuss the optimization problem setting which embraces all the formulations we met before in Sect \ref{4Set}. 
We no longer treat the infinity point both in the domain of definition and the range of rational function as an exceptional one. 
Real line extended by a point at infinity becomes a \emph{real projective line} $\mathbb{R}P^1:=\hat{\mathbb{R}}=\mathbb{R}\cup\{\infty\}$
which is a topological circle. We consider two collections of disjoint closed segments on 
the extended real line: $E$ consisting of $m\ge 2$ segments and $F$ of just two segments. The segments of both $E$ and $F$ are of two types:
$E:=E^+\sqcup E^-$; $F:=F^+\sqcup F^-$. Neither the stopbands $E^-$ nor the passbands $E^+$ are empty sets.

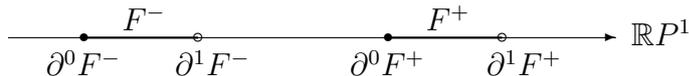
\begin{figure}
\begin{picture}(160,22)(-35,0)
\put(5,10){\vector(1,0){80}}
\put(87,9){$\mathbb{R}P^1$}
\put(15,10){\circle*{1}}
\put(10,5){$\partial^0F^-$}
\put(30,10){\circle{1}}
\put(27,5){$\partial^1F^-$}
\put(55,10){\circle*{1}}
\put(50,5){$\partial^0F^+$}
\put(70,10){\circle{1}}
\put(68,5){$\partial^1F^+$}
\put(20,11){$F^-$}
\put(60,11){$F^+$}
\thicklines
\put(15,10){\line(1,0){15}}
\put(55,10){\line(1,0){15}}
\end{picture}
\caption{The ordering of four endpoints of $F$}
\label{dF}
\end{figure}

\subsection{Large classes}
Consider the space $\mathbb{R}_n(x)\subset\mathbb{R}(x)$ of degree not greater than $n$ rational functions
which we treat as the the mappings from one copy of real projective line to the other. 
Unless otherwise is explicitly indicated, in this space we use the topology induced by the coefficients of rational functions. 
Namely, we treat the homogeneous coordinates in the real projective space as suitably ordered coefficients of a rational function 
and obtain a natural map $\mathbb{R}P^{2n+1}\to\mathbb{R}_n(x)$ which is $1-1$ almost everywhere. 
In a natural way it transfers \cite{Kelly} the topology from the projective space to the space of rational functions. 
Along with the topology of coefficients one can consider other topologies 
like uniform convergence on the whole projective line of arguments (or its parts) with respect to Fubini-Study metrics on 
the target projective line. All reasonable topologies on the space  $\mathbb{R}_n(x)$ coincide outside the variety 
$\mathbb{R}_{n-1}(x)$ of functions with
the degree strictly lower than $n$. In the vicinity of mentioned discriminant variety (described as zero of appropriate resultant) we may observe the differences: if a real zero and a pole of a rational function coalesce, the convergence  to the limit function of lower degree takes place in the sense of coefficients, but it is not uniform.

\begin{dfn}
Large class  ${\cal R}_n(E,F)$ consists of the fixed degree
$\deg R=n$ real rational functions $R(x)$  with the property 
\be 
R(E^+)\subseteq F^+ \quad and \quad R(E^-)\subseteq F^-.
\label{REinF}
\ee
The closure (in the topology induced by the coefficients) of the large class  is the 
set of bounded degree $\deg R\le n$ rational functions with the same property \eqref{REinF} and denoted as $\overline{{\cal R}_n(E,F)}$. 
\end{dfn}

We use the adjective ``large'' for the description of the functional class since it usually has several components  
-- called the small or Stiefel classes,  which is explained in the next section.

\subsection{Topological indexes and  small classes}
Separated values at pass and stopbands produce a decomposition of the set of rational functions ${\cal R}_n(E,F)$ 
into pieces with different topological properties. This phenomenon was first observed for the problem setting \ref{minDev}, seemingly
by E.Stiefel \cite{Sti} in 1961 and elaborated in detail by R.A.-R. Amer in his PhD thesis \cite{AS} in 1964. Indeed, once the value of the goal function from \ref{minDev} is finite, a rational function $R(x)$ has no 
zeros on $E^-$ and no poles on $E^+$, hence on each band from $E$  
the polynomial in  either the numerator or the denominator of the fraction  of $R(x)$ has a definite sign. This array of signs is a topological invariant. Here we reintroduce the decomposition into topological classes for the general projectively invariant setting.

Identification of the opposite points of a circle $S^1$ gives a double cover $S^1\to\mathbb{R}P^1$.
Trying to lift the mapping $R(x):\quad \mathbb{R}P^1\to\mathbb{R}P^1$ to the double cover of the target space $\tilde{R}(x):\quad \mathbb{R}P^1\to S^1$, we encounter  a topological obstruction to the existence of $\tilde{R}$, namely the mapping degree or the winding number of $R(x)$ modulo 2. A simple calculation shows that this value is equal to the algebraic degree $\deg R$ $mod~2$. However, the lift exists  on any simply connected piece of $\mathbb{R}P^1$. Consider any \emph{transition band}  
$T$, a component of the complement of $E$ to the projective line.
The set $F\subset\mathbb{R}P^1$ lifted to the circle $S^1$ consists of four segments, two consecutive of which we call $\tilde{F}_0$
and another two as $\tilde{F}_1$. Each of two antipodal sets $\tilde{F}_0$, $\tilde{F}_1$ descends  1-1 back to $F$. The mapping $R(x):\quad T\to\mathbb{R}P^1$ has two lifts $\tilde{R}$ to the covering circle $S^1$ and exactly one of them has value $\tilde{R}(x)\in \tilde{F}_0$ on the left endpoint $x=\partial^0T$ of $T$. On the opposite side of the interval $T$ the same function $\tilde{R}(x)$ takes value in the set $\tilde{F}_\sigma$ with well defined $\sigma=\sigma(R,T)\in\{0,1\}$ which we call the \emph{transition index} of the band $T$. Totally, the function $R(x)$ defines a locally constant map ${\bm\sigma}:=\{\sigma(R,T)\}_T$: $\mathbb{R}P^1\setminus E\to$ $\mathbb{Z}_2$ with the constraint:
\be
\label{constraint}
\sum_{T} \sigma(R,T) =\deg~R ~~mod ~2. 
\ee

\begin{figure}
\includegraphics[width=.5\textwidth]{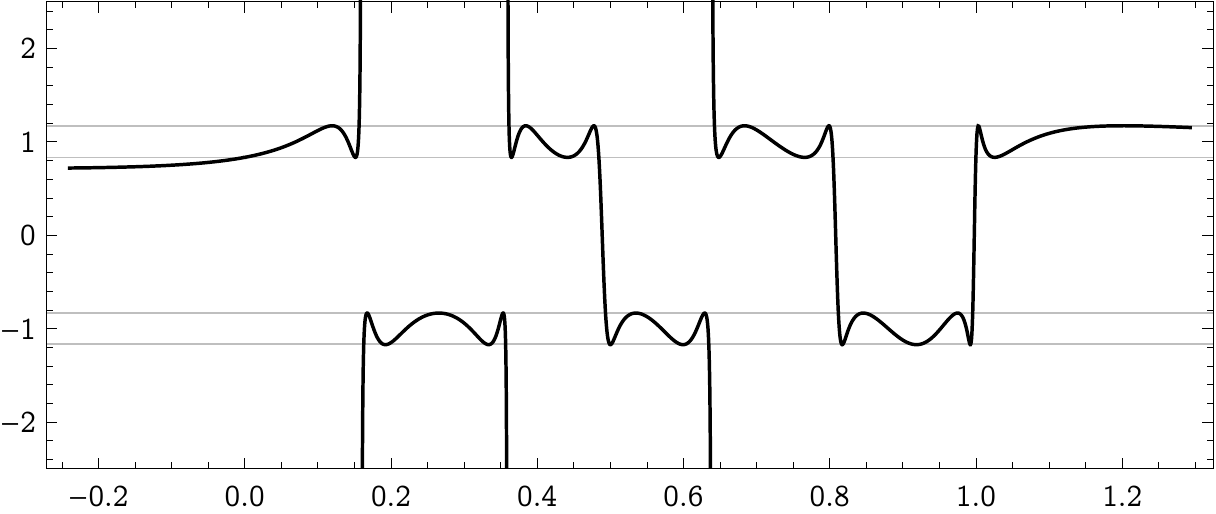}
\includegraphics[width=.5\textwidth]{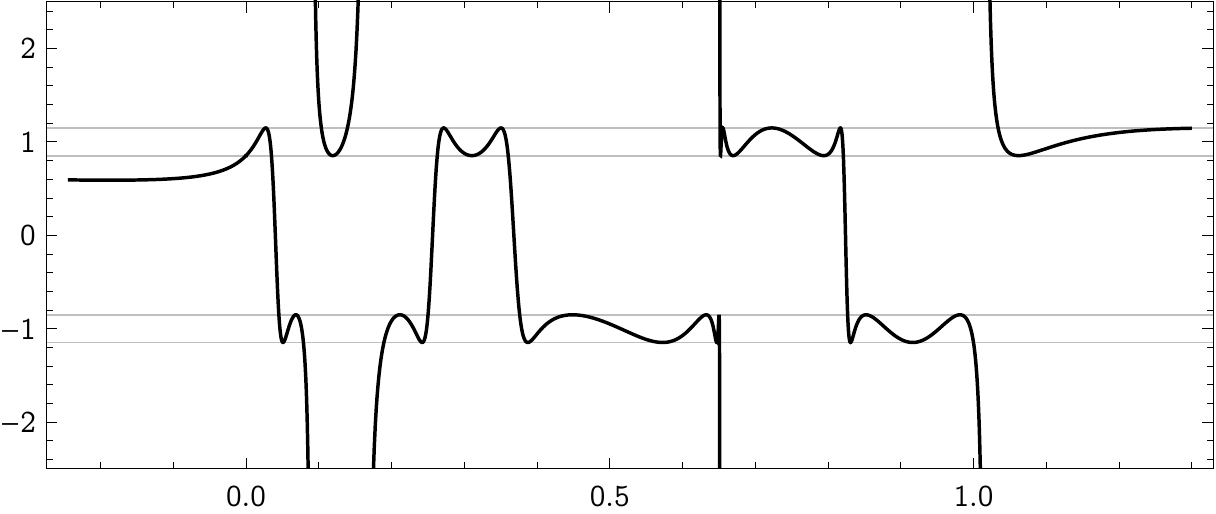}
\caption{Solutions with $F^\pm$ being the vicinities of $\pm1$
and different sets of transition indexes. Computed by S.Lyamaev.} 
\end{figure}

A practical way of computing the topological indexes is the following. Consider the  
universal covering $\mathbb{R}\to$ $\mathbb{R}P^1=\mathbb{R}/\pi\mathbb{Z}$ of the kind
$\mathbb{R}\ni\phi\to (\sin(\phi):\cos(\phi))\in\mathbb{R}P^1$. Its local inversion is given by 
$\phi(P:Q)=Arg(Q+iP) ~~(+\pi\mathbb{Z})$
for homogeneous coordinates $(P:Q)$ of the projective line. Fix a union of two segments $\tilde{F}_0\subset\mathbb{R}$ 
which are mapped 1-1 to the set $F\subset\mathbb{R}P^1$ under the universal covering. Then the full preimage of $F$ will have the appearance
$\tilde{F}_0+\pi\mathbb{Z}$. Representing an element $R(x)$ of ${\cal R}_n(E,F)$ as the irreducible ratio of  real polynomials
$P(x):Q(x)$ we can lift the rational function to the universal cover of the target space:
$$
\tilde{R}(x)=Arg(Q(x)+iP(x))
$$
which is normalized by the condition $\tilde{R}(\partial^0T)\in\tilde{F}_0$ (here we possibly have to change signs of both polynomials $P,Q$). 
The value of $\tilde{R}$ on the opposite side $\partial^1T$ of the segment  lies in $\tilde{F}_0+\pi\sigma$ with uniquely defined integer $\sigma=\sigma(R,T)$. Of course, this index depends on the choice of the auxiliary set $\tilde{F}_0$, but this dependency is well controlled. Introduced above binary index in what follows we mostly will be interested in  is merely a reduction of the integer index $\sigma$ mod 2.

The above procedure defines a transition index (integer or binary) for any interval $G$ of the projective line, with endpoints in $E$. 
Simple calculation shows that it is determined by the indexes of transition bands contained in $G$: 
$$\sigma(R,G)=\sum_{T\subseteq G}\sigma(R,T).$$

Transition indexes $\sigma(R,T)$ do not change with the increase of $F$ which induces simultaneous deformation of the auxiliary covering set $\tilde{F}_0$. Transition indexes are topological invariants of the rational function $R$ that is remain intact once we continuously deform the function within the class ${\cal R}_n(E,F)$ (in the usual sense, $R(x,s)$ is a continuous map from $T\times [0,1]\to \mathbb{R}P^1$, 
rational in the first argument). Indexes may change only iff zeros and poles of $R$ merge on some transition band in the process of deformation which inevitably brings us to the fall of the degree which shows a simple proposition:

\begin{lmm}
Let $R(x)\in {\cal R}_{n}(E,F)$, $0\in T$ and $R(0)\neq 0,\infty$,  then for small real $\epsilon\neq0$ 
the integer indexes $\sigma$ are related as follows: 
\be\label{indexdrop}
\sigma(R(x)\frac{x+\epsilon}x,T)=\sigma(R,T)-\sign(\epsilon R(0)),
\ee
here $\sign(\cdot)=\pm1$.
\end{lmm}
Proof: direct calculation. ~~\bl

\begin{dfn}
The elements of ${\cal R}_n(E,F)$ with the fixed array ${\bm\sigma}:=\{\sigma(T)\}_T$ of binary  transition indexes 
make up a subset $ {\cal R}_n(E,F,{\bm\sigma})$ we call the small or Stiefel class.
\end{dfn}
Some properties of the introduced small classes are described in:

\begin{lmm}
1. Double monotonicity:
$$
{\cal R}_n(E,F,{\bm\sigma})\subseteq {\cal R}_n(E',F',{\bm\sigma}), \qquad  F\subseteq F' ~and~ E\supseteq E',
$$
provided each segment of $E$ contains a unique segment of $E'$, which is of the same $\pm$-type.

2. Double projective invariance
$$
\beta\circ {\cal R}_n(E,F,{\bm\sigma})\circ\alpha^{-1}={\cal R}_n(\alpha E, \beta F, {\bm\sigma'}), \qquad \alpha,\beta\in PGL_2(\mathbb{R}),
$$
here $\sigma'(\alpha T)=\sigma(T)$ for all transition bands $T\subset\mathbb{R}P^1\setminus E$
and specified below consistent choice of the auxiliary set $\tilde{F}_0$ 
for the definition of transition indexes.\\

3. Class ${\cal R}_n(E,F,{\bm\sigma})$ is empty when $n<\sharp\{T:\quad \sigma(T)=1\}$.\\

4. Small classes may be (and usually are) not connected. 
\end{lmm}

Proof. 1) For a given rational function $R$ the growth of the set $F$ and shrinking of the set $E$ does not spoil the restriction \eqref{REinF}
and does not change any transition index $\sigma(R,T)$. 

2) The first assertion $\beta\circ {\cal R}_n(E,F)\circ\alpha^{-1}={\cal R}_n(\alpha E, \beta F)$
follows immediately from the definition of large classes. It remains to check the transition indexes.

Let us clarify the dependence of the binary index $\sigma(T)$ on the choice of the auxiliary set $\tilde{F}_0$.
The full preimage of $F=F^+\sqcup F^-$ under the double cover consists of four segments cyclically labeled e.g. as
$\tilde{F}^+_0,\tilde{F}^-_0,\tilde{F}^+_1,\tilde{F}^-_1$. Other decompositions of the full preimage arise after the relabeling $\tilde{F}^+_0\leftrightarrow\tilde{F}^+_1$ or $\tilde{F}^-_0\leftrightarrow\tilde{F}^-_1$ or both applied to the initial labeling. 
For the first and the fourth labeling the binary index is the same while for the second and the 
third case the index $\sigma(T)$ reverses (compared to the first labeling) exactly when the transition band $T$ is surrounded by stop and pass bands.

The automorphism $\beta$ of the projective line may be lifted in two ways to the automorphism $\tilde{\beta}$ 
of the doubly covering circle. Those two lifts differ by the (pre- or post-) composition with the antipodal map and we fix one of them.
Now we choose the agreed labeling for the auxiliary sets $\tilde{F}_0$ and $\tilde{(\beta F)}_0:=\tilde{\beta}\tilde{F}_0$ 
which brings us to the  above relation for the binary transition indexes.  Indeed, one of the lifts for the map $\beta\circ R\circ\alpha^{-1}$ is $\tilde{\beta}\circ\tilde{R}\circ\alpha^{-1}$ as we see it from the diagram \eqref{LiftsDiagram} where vertical arrows denote double coverings:
\be
\begin{array}{ccccccc}
&&&& S^1 & \overset{\tilde{\beta}}{\rightarrow} & S^1\\
&&& \overset{\tilde{R}}{\nearrow} & \downarrow && \downarrow\\
\alpha T & \overset{\alpha^{-1}}{\rightarrow} & T & \overset{R}{\rightarrow} & \mathbb{R}P^1 & \overset{\beta}{\rightarrow} & \mathbb{R}P^1\\
\end{array}
\label{LiftsDiagram}
\ee

We easily check that on the endpoints of the segment
$\alpha T$ the lifted map takes the values from $\tilde{(\beta F)}_0$ and $\tilde{(\beta F)}_\sigma$, with $\sigma=\sigma(R,T)$.
The order of values depends on the orientation of $\alpha$, but in any case $\sigma(\beta\circ R\circ\alpha^{-1},\alpha T)=\sigma(R,T)$.

3) Suppose the converse is true and there is a degree $n$ rational function $R$ whose binary transition index $\sigma$
takes the value $1$ at more than $n$ transition bands $T$. The auxiliary sets $\tilde{F}_0$ and $\tilde{F}_1$ lie in the opposite semicircles of 
$S^1$. The endpoints of this semicircle are glued to a point of real projective line which becomes a value of $R(x)$ 
more than $n$ times which is impossible. If each transition band  is surrounded by a stop and a passband (as it usually happens in practice),
then possibly relabeling the components of the total lift of the set $F$ (see section 2 of this proof) we arrive at the stronger version  of 
the small class vanishing criterion: $n<\max_{j=0,1}\sharp\{T:\quad \sigma(T)=j\}$. This happens e.g.  when $2n$ is strictly less than the number of components in $E$.

4) Take any function $R(x)$ of the class ${\cal R}_{n-1}(E,F)$, we assume w.l.o.g. that $x=0$ is contained in the transition band $T$
and $R(0)\neq0,\infty$. For small real $\epsilon$ two functions $\frac{x\pm\epsilon}xR(x)$ belong to the same large class 
${\cal R}_n(E,F^*)$ with $F^*$ slightly larger than $F$. From Lemma \ref{indexdrop} it follows that two functions 
have different integer transition indexes but the same binary index at $T$, therefore they belong to the different components of the same (binary) small class ${\cal R}_n(E,F^*,{\bm\sigma})$.
~~~~\bl


\subsection{Cross ratio and extremal problem}
The set of values $F$ modulo \emph{projective (=linear-fractional) transformations} depends on a single value -- 
cross ratio of its four endpoints. Suppose the endpoints $\partial F$ are cyclically ordered as follows 
$\partial^0F^-$, $\partial^1F^-$, $\partial^0F^+$, $\partial^1F^+$ -- see Fig. \ref{dF} 
then the cross ratio of four endpoints we define as follows
\begin{dfn}
$$
\kappa(F):=\frac{\partial^1F^+-\partial^1F^-}{\partial^1F^+-\partial^0F^+}:
\frac{\partial^0F^--\partial^1F^-}{\partial^0F^--\partial^0F^+}>1.
$$
\end{dfn}

Cross ratio depends on the order of four participating endpoints and may take six values interchanged by the elements of the  unharmonic group, however it survives under the action of Klein's quadratic group which is one of the assertions of 
\begin{lmm}
1) Definition of $\kappa(F)$ is independent of the projective line orientation (i.e. relabeling $0\leftrightarrow1$)
and changing the type of components of $F$ (i.e. interchange of indexes $\pm$).\\
2) $\kappa(F)$ decreases with the growth of its argument: if $F'\subset F$ then $\kappa(F')>\kappa(F)$.\\
\end{lmm}

Now we can give a projectively consistent formulation of the optimization problem for electrical filter.  
The idea behind this optimization is the following: as we squeeze the set of values $F$, the  nonempty functional class 
${\cal R}_n(E,F,{\bm\sigma})$ shrinks and we have to catch the moment -- quantitatively described by the cross ratio $\kappa(F)$ -- when the class disappears.

{\bf Problem.}
Given the set of bands $E$ and the transition index array $\bm\sigma$ consistent (see \eqref{constraint}) with the fixed degree $n$, find
\be
\label{minsigma}
\varkappa(n,E,{\bm\sigma}):=\inf\{\kappa(F): \quad {\cal R}_n(E,F,{\bm\sigma})=\emptyset\}. 
\ee

\begin{rmk}
1) Same problem may be formulated for the large classes without fixing the array of transition indexes. Its solution is the maximal value of  
$\varkappa(n,E,{\bm\sigma})$ over the admissible arrays $\bm\sigma$. \\
2) In problem formulation \ref{Zolo3} the set $F^+=[-\theta,\theta]$ and the set $F^-=[1/\theta,-1/\theta]$; $\kappa(F)=\left(\frac12(\theta+1/\theta)\right)^2$. In setting \ref{Zolo4} the sets $F^\pm=\pm[1-\mu,1+\mu]$ and $\kappa(F)=\mu^{-2}$. 
\end{rmk}

The set $F$ with the extremal value of $\kappa>1$ may correspond to the empty small class ${\cal R}_n(E,F,{\bm\sigma})$.
To take into account solutions with defect (= of  degree less than $n$),  we introduce the notion of the limit class.

\begin{dfn}
The   limit class we call the intersection of nested closed sets
\be\label{ExtCl}
\overline{{\cal R}_n}(E,F,{\bm\sigma}):=\bigcap_{Int~F'\supset F} \overline{{\cal R}_n(E,F',{\bm\sigma})}.
\ee
\end{dfn}

Let us introduce several characteristics of a rational function $R$ lying in the closed  class $\overline{{\cal R}_n(E,F)}$.
The set of extremal points $Ext(R)$ consists of points $e\in E$ mapped to the boundary of $F:$ ~$R(e)\in \partial F$. Each extremal point $e$ has a parity $\epsilon(e)=0/1$
depending to which boundary component $\partial^0F/\partial^1F$ (=left/right) does the value $R(e)$ belong to. 
The complement $\mathbb{R}P^1\setminus Ext(R)$
to the set of extremal points is a disjoint union of open intervals $G$ (gaps) which fall into two types: for even $G$ its endpoints 
have the same parity $\epsilon$, for odd gap the endpoints are of different parities. Now for each gap $G$ we introduce the \emph{differential transition index} of  $R$ as the sum over all transition bands $T$ contained in the gap $G$ of the differences of transition indexes of $R$ and the given ${\bm\sigma}=\{\sigma(T)\}_T$, all taken modulo 2:
\be
\label{sigmaG}
\delta\sigma(G,R):=(\sum_{T\in G} \sigma(R, T)-\sigma(T))_2=(\sigma(R, G)-\sigma(G))_2 ~~~\in\{0,1\},
\ee
here and in what follows $(\cdot)_2$ is the reduction of integers modulo 2 with the values in the set $\{0,1\}$.
We distinguish between global differences which occur on two types of gaps:
\be
\label{globalS}
\Sigma^0(R):=\sum_{even~G}\delta\sigma(R,G);\qquad
\Sigma^1(R):=\sum_{odd~G}\delta\sigma(R,G).
\ee
here the first sum is taken over all even gaps $G$ (endpoints of the same parity) and the other sum is taken over 
odd gaps (endpoints of different parity).

We note two simple facts for the defect of the functions at the boundary of the closed small class.
\begin{lmm}
Let  $R\in\overline{{\cal R}_n(E,F,{\bm\sigma})}$ and 
$\Sigma(R):=\sum_T(\sigma(R,T)-\sigma(T))_2$ be the Hamming distance of two binary arrays: 
actual transition indexes of $R$ and those inherited from the small class. Then\\
1.
\be
\label{defect}
\Sigma(R)\le d(R):= n-\deg(R),
\ee
2.
\be
\label{dparity}
d+\Sigma^0+\Sigma^1\in2\mathbb{Z}, \qquad\Sigma(R)+d\in2\mathbb{Z}.
\ee
\end{lmm}
Proof. The defect $d$ is the number of cancellations of (simple) zeros/poles of a rational function
when approaching the boundary of the small class.
The binary index $\sigma(R,T)$ inverses exactly when there are odd number of cancellations at the 
transition band $T$. ~~\bl

\subsection{Equiripple property}
\begin{dfn}
We say that cyclically ordered extremal points of the function $R(x)\in\overline{ {\cal R}_n(E,F)}$  
make up an alternation set iff any two consecutive points have opposite parity (in particular, their total number is even).
The maximal number $Alt(R)$ of alternation points  is equal to the number of odd gaps $G\subset\mathbb{R}P^1\setminus Ext(R)$.
\end{dfn}

\begin{thrm}
\label{main}
If the value $\varkappa(n,E,{\bm\sigma})>1$, then each extremal limit class $\overline{{\cal R}_n}(E,F,{\bm\sigma})$
with $\kappa(F)=\varkappa(n,E,{\bm\sigma})$ contains a unique function 
$R(x)$. This function is completely characterized by the property 
\be
\label{Alt}
Alt(R)\ge n+2+\deg R-\Sigma^0(R)+\Sigma^1(R).
\ee
Note that both sides of the inequality are even.
\end{thrm}
Proof of this theorem is based on two technical propositions.

\begin{lmm}
\label{euclid} (Achiezer, \cite{Ach2})
Let $P$ and $Q\in\mathbb{C}[x]$ be two mutually prime polynomials of the same degree $n$ and $S$ be an arbitrary polynomial of degree $m$, then 
the equation 
\be
\label{Euclid}
S=pQ-qP
\ee
admits polynomial solutions $p,q$ of degree $\le \max(m-n, n-1)$.
\end{lmm}
Proof. Consider two Lagrange polynomials   $q_1$ and $p_1$  which interpolate $S/P$ at the zeros of $Q$ and $S/Q$ at the zeros of $P$ respectively, $\deg~ p_1, q_1\le n-1$. In the case of multiple zeros we require not only the equality of values of polynomials but also the equality of several first derivatives. Now we define the polynomial $2r:=(S-p_1Q-q_1P)/(PQ)$ which is identically equal to zero when $m<2n$ or $\deg~ r=m-2n$
when $m\ge 2n$. One easily checks that $p:=p_1+rP$; $q:=-q_1-rQ$ is a solution of \eqref{Euclid} and any other solution has the appearance 
$p+r_1P$; $q+r_1Q$ with arbitrary $r_1\in\mathbb{C}[x]$. 
~~\bl

\begin{lmm}
\label{zeros}
Let $R(x)$ and $R'(x)$ belong to the same closed class $\overline{{\cal R}_n(E,F)}$ and 
$e_1$, $e_2$ be two extremal points of $R(x)$.
Suppose the values of $R$ and $R'$ do not coincide at each of $e_j$, then the number of 
points $t\in[e_1,e_2]$ such that $R(t)=R'(t)$ counted with the multiplicity of tangency has the same parity as
\be
\label{parity}
\epsilon(e_1)+
\epsilon(e_2)+
\sum\limits_{e_1<T<e_2} (\sigma(R,T)+\sigma(R',T)),
\ee
where summation is taken over all transition bands in the segment $[e_1,e_2]$ of oriented projective line.
\end{lmm} 

\begin{rmk}
Multiplicity of coincidence of functions $R$, $R'$ at a point $t$ is equal to the order of zero of their difference
at this point provided $R(t)=R'(t)\neq\infty$. Otherwise we use another projective coordinate for the target projective line
and apply the previous definition. In other words, the multiplicity of tangency at a point is the multiplicity of zero of the 
determinant $PQ'-QP'$ where $P/Q:=R$ and $P'/Q':=R'$ are irreducible representations of rational functions as ratios of polynomials.
\end{rmk}
\begin{rmk}
The coincidence multiplicity is odd exactly when the difference $R-R'$ changes its sign in the small vicinity of the 
coincidence point.
\end{rmk}
\begin{rmk}
Here we do not exclude the case $e_1=e_2$, the segment between those points then  means either 
one point or the whole projective line. 
\end{rmk}

Proof. We use identifications $S^1=\mathbb{R}/2\pi\mathbb{Z}$ and  $\hat{\mathbb{R}}=\mathbb{R}P^1=\mathbb{R}/\pi\mathbb{Z}$
which are implemented by the functions $\exp(i\phi)$ and $\tan(\phi)$ respectively. A lift of the mapping 
$R(t):[e_1,e_2]\to\hat{\mathbb{R}}$ to the double cover of the target space is implemented now by the formula
$\tilde{R}(t)=\phi(t)=\Arg(Q(t)+iP(t))\quad(mod~2\pi)$ where $R(t)=P(t)/Q(t)$ is the irreducible representation of the rational function.
The choice of the lift is twofold and can always be made so that $\tilde{R}(e_1)\in\tilde{F}_0$. The values of this lift at the endpoints of the segment $[e_1,e_2]$ are as follows:


\be
\tilde{R}(e_1)=\partial^{\epsilon_1}\tilde{F}_0^{*}
\quad \mod~2\pi,
\ee
\be
\tilde{R}'(e_1)\in[\partial^0\tilde{F}_0^{*},\partial^1\tilde{F}_0^{*}]\setminus\partial^{\epsilon_1}\tilde{F}_0^{*}
\quad \mod~2\pi,
\ee
where the sign $*=\pm$ is defined by the type of the band $E^\pm\ni e_1$ and $\epsilon_1:=\epsilon(e_1)$;
\be
\tilde{R}(e_2)=\partial^{\epsilon_2}\tilde{F}_0^{**}+\pi\sigma(R,[e_1,e_2]),
\qquad \mod~2\pi
\ee
\be
\tilde{R}'(e_2)\in[\partial^0\tilde{F}_0^{**},\partial^1\tilde{F}_0^{**}]\setminus\partial^{\epsilon_2}\tilde{F}_0^{**}+
\pi\sigma(R',[e_1,e_2]), \qquad\mod ~2\pi,
\ee
where the sign $**=\pm$ is defined by the type of the band $E^\pm\ni e_2$, $\sigma(R,[e_1,e_2])=\sum\limits_{e_1<T<e_2} \sigma(R,T)$, $\epsilon_2:=\epsilon(e_2)$.

Consider the difference of the lifted values $\tilde{R}=\phi,\tilde{R}'=\phi'$ at the endpoints:
\be
\phi'(e_1)-\phi(e_1)
\quad\in(-1)^{\epsilon_1}(0,|F^*|] 
\quad \subset(-1)^{\epsilon_1}(0,\pi) 
~~ \mod~2\pi,
\label{de1} 
\ee
$$
\phi'(e_2)-\phi(e_2)
\quad\in(-1)^{\epsilon_2}(0,|F^{**}|]
+\pi(\sigma(R',[e_1,e_2])-\sigma(R,[e_1,e_2]))\subset
$$

\be
\label{de2} 
(-1)^{\epsilon_2}(0,\pi)+\pi(\sigma(R',[e_1,e_2])-\sigma(R,[e_1,e_2])) 
\quad \mod~2\pi,
\ee
here $|F^\pm|<\pi$ is the length of the component of $F$ in Fubini-Study metric on the projective line.
The coincidence of the values of $R,R'$ at a point is equivalent to $\phi'-\phi\in\pi\mathbb{Z}$ at the point.
The multiplicity of the coincidence $R=R'$ is equal to the multiplicity of zero of the value:
$$
(\phi'(t)-\phi(t))\mod\pi\sim \sin(\phi'(t)-\phi(t)) =Im~\exp(i\phi'(t)-i\phi(t))=
$$
$$
Im~
\frac{Q-iP}{\sqrt{P^2+Q^2}}\frac{Q'+iP'}{\sqrt{P'^2+Q'^2}}\sim(PQ'-QP')(t).
$$ 

Finally we see it from the formulas \eqref{de1} and \eqref{de2} that the number of transitions of the value $\delta\tilde{R}$ on the segment $[e_1,e_2]$ from one semicircle $\pm(0,\pi)$ to the other is equal to the value \eqref{parity} modulo 2. ~~\bl

\section{Main theorem proof} 

{\bf 0. Existence.}
We show that the closed sets in the definition \eqref{ExtCl} of the limit extremal class make up a centered family, that is intersection of 
finitely many of them is nonempty. Using the compactness of the ambient set of 
bounded degree rational functions we easily get that all sets of the family have a common point \cite{Kelly}.

Indeed, any finite intersection of the small classes closures $\overline{{\cal R}_n(E,F',{\bm\sigma})}$
contains at least the small class ${\cal R}_n(E,F'',{\bm\sigma})$ with $F''$ being 
the finite intersection of the sets $F'$. Since $Int~F'\supset F$, $\kappa(F'')<\kappa(F)$ and the latter small class is not empty
because $\kappa(F)>1$ is the extremal value.

{\bf 1. Sufficiency.}
Suppose that $R(x)$ from the closed class $\overline{{\cal R}_n(E,F)}$ has big enough 
(specified in the formulation of Theorem \ref{main}) alternation set 
defined with respect to $F$ and the given binary array $\bm\sigma$. 
We show that $\kappa(F)$ is extremal value in \eqref{minsigma} 
and the limit class $\overline{{\cal R}_n}(E,F,{\bm\sigma})$ contains the unique point $R$. 
Let otherwise $R'$ be any function in the limit set $\overline{{\cal R}_n}(E,F',{\bm\sigma})$ with $F'\subseteq F$. 

The complement of the set $Ext(R)$ to the projective line consists of finitely many opened intervals (gaps)
$G$. Consider a chain of $k\ge1$ consecutive gaps   
such that $R(e)=R'(e)$ for any interior extremal point $e\in Ext(R)$ of the cluster whereas the values of $R,R'$ 
are different at both endpoints $e_1$, $e_2$ of the chain. Let us assess the number of coincidences (counting the multiplicities) of $R,R'$  inside the cluster of gaps. This number is at least $k-1$, however Lemma \ref{zeros}  can guarantee an extra coincidence point at $[e_1,e_2]$ once the value \eqref{parity} calculated for the cluster has the same parity as $k$. Let the chain 
contain $m_0$ even and $m_1$ odd gaps $G$, $m_0+m_1=k$, then the sum of parities of the endpoints of cluster is $m_1$ modulo 2. 
The difference of transition indexes for the functions $R$, $R'$ over all transition bands in the cluster amounts to 
$\sum_{G\subset[e_1,e_2]}\delta\sigma(G)+\delta\sigma'(G) ~mod~2$, where we for brevity designated 
$
\sigma(G):=\sigma(G,R); ~\sigma'(G):=\sigma(G,R')
$
and the differential transition indexes are calculated with respect to the array $\bm\sigma$ of binary indexes
inherited from the optimization problem. Hence the number of coincidence points in the cluster is at least 
$$
k-(k+m_1+\sum_{G\subset[e_1,e_2]}\delta\sigma(G)+\delta\sigma'(G))_2\ge
$$
$$
m_1+\sum_{even~G} (\delta\sigma(G))_2-\sum_{odd~G} (\delta\sigma(G))_2-\sum_G (\delta\sigma'(G))_2.
$$
To prove the last inequality we note that its left hand side is not less than $k-1$, whereas the 
right hand side is not greater than $k-1$ with one exception: $(\delta\sigma'(G))_2=0$ for all gaps $G$, 
$(\delta\sigma(G))_2=0$ for all odd gaps and $(\delta\sigma(G))_2=1$ for all even gaps. In the latter case
we have the equality of both sides.

Summing up the latter estimations over all clusters we get a lower bound for the number of 
the coincidence points of two rational functions  $R$ and $R'$ on the projective line. Taking into account the 
inequality \eqref{Alt} for the number of cyclic alternation points and the assessment \eqref{defect} for the defect of $R'$, 
the lower bound we consider becomes :
\be
\label{clue}
Alt(R)+\Sigma^0(R)-\Sigma^1(R)-\Sigma(R')>\deg(R)+\deg(R'), 
\ee
where from it follows that $R=R'$ identically. The omitted case when $R=R'$ in each extremal point of $R$
immediately brings us to the assessment in the left hand side of \eqref{clue} for the number of coincidence points 
since the number of even gaps $G$ is not less than $\Sigma^0(R)$.

{\bf 2. Necessity.} Suppose a function $R\in$ $\overline{{\cal R}_n}(E,F,{\bm\sigma})$
possesses not enough number of alternation points. We explicitly indicate a function $R'$ in the class
${\cal R}_n(E,F',{\bm\sigma})$ with  $F'\subset F$. To avoid technical difficulties we
fix the coordinate in the target projective line so that $\infty\not\in F$
and $R(\infty)\not\in0\cup\infty\cup\partial F$. This is particular means that fraction $R=P/Q$ is 
the ratio of mutually prime polynomials of the same degree $n-d$, where $d\ge 0$
is the defect of $R$. 

We are looking for a modification of $R$ among functions of the kind
\be
\label{modified}
R'(x):=\frac{M(x)\prod_{s=1}^\Sigma(x-x_s)P(x)-\tau p(x)}{M(x)\prod_{s=1}^\Sigma(x-x_s)Q(x)-\tau q(x)},
\ee
where $x_s$ is any point in the transition band $T_s$ with the inverted binary index $\sigma(R,T_s)\neq\sigma(T_s)$
modulo 2 and $\Sigma=\Sigma(R)$ is the total number of such bands. Polynomial $M(x)$ of even degree $d-\Sigma$ is positive on the real axis;  
$\tau $ is a small real parameter.  Polynomials $p(x)$ and $q(x)$ of degrees not greater than $n$  determine the direction 
of deformation in the space of rational functions and will be chosen later. The fraction $R'$ belongs to some class $R_n(E, F',{\bm\sigma})$ 
once its numerator and denominator have no common factors. The set $F'$ here will be strictly smaller than $F$ for small values 
of $\tau$ if the difference $\delta R(x):=R'(x)-R(x)$ has the sign $(-1)^{\epsilon(x)}$ in each extremal point $x\in Ext(R)$  -- see e.g.\cite{Ach2} Chap. 2 or \cite{Akh1} Chap. 9, for more details\footnote{Once $Ext(R)$ is empty, the set $F$ surely may be squeezed.}. The sign of the difference $\delta R(x)$, $x\in E$, for small $\tau$ is the same as the sign of the polynomial $\tau\prod_{s=1}^\Sigma(x-x_s)(P(x)q(x)-Q(x)p(x))=:\tau L(x)$. We can explicitly construct the latter polynomial $L$ provided the number of alternation points of $R$ is insufficient.

Polynomial $L$ should have even/odd number of zeros in each even/odd finite gap $G$. It already has $\Sigma$ zeros $x_s$
used to recover the array $\bm\sigma$ of binary transition indexes. So we place 
additional zero to each gap $G$ where the parity of $\delta\sigma(R,G)$ is opposite to the parity of the gap itself.
We get a polynomial $L$ of degree $Alt(R)-\Sigma^1+\Sigma^0+\Sigma$ or one less as we need not place an additional zero to the gap at infinity. 
Canceling $L(x)$ by $\prod_{s=1}^\Sigma(x-x_s)$ we obtain a left hand side $S(x)$ of the functional equation \eqref{Euclid} for polynomials $p,q$. The latter according to Lemma \ref{euclid} always has a solution for mutually prime $P,Q$. The solution has degrees
of $p,q$ strictly greater than $n$ exactly when $\deg L-\Sigma-n+d>n$. Taking into account the parities of $Alt(R)$ and that of the  defect 
\eqref{dparity} we obtain the required in Theorem \ref{main} number of cyclic alternation points for $R$. 

It remains to check that fraction $R'(x)$ has no cancellations for small $\tau$. We assume w.l.o.g. that only simple zeros 
of numerator and denominator cancel when $\tau=0$. Those zeros smoothly depend on time $\tau$ and have finite velocities at the initial moment
which may be explicitly calculated. The velocity of each of those zeros in numerator is different from that for denominator when $\tau=0$,
for otherwise the polynomial $L(x)M(x)$ would have multiple zeros. We have enough freedom in the choice of the latter polynomial
to avoid this occasion.  ~~~\bl

\section{Conclusion}
We have elaborated a universal setting for the optimization problem of multiband electrical filter which apparently 
embraces all previously known settings. 
Novel viewpoint at this problem is consistent with the  important property of the latter, its projective invariance.
A notion of cyclic alternation points is introduced to give the optimality criterion for the new setting.
This criterion also involves the information about decomposition of the space of rational functions into topological classes which 
guarantee the uniqueness of the solution. Extremal functions introduced in \cite{B10} give us examples with nontrivial additional terms $\Sigma^0$, $\Sigma^1$ and will be described in a separate publication. Omitted technical (but very important in the engineering practice) details like the weight of the approximation or possible restrictions on participating rational functions can be also incorporated to the elaborated scheme.

\begin{verbatim}
Institute for Numerical Math., Russian Academy of Sciences, 
Address: Russia 119991 Moscow ul. Gubkina, 8   
Email: ab.bogatyrev@gmail.com, gourmet@inm.ras.ru 
\end{verbatim}

\end{document}